\documentclass[prl,superscriptaddress,showpacs,twocolumn]{revtex4}
\usepackage{graphicx,amsfonts}
\def\figwidth{8cm}
\begin{document}
\title{Specific heat of classical disordered elastic systems}
\author{Gregory Schehr}
\affiliation{CNRS-Laboratoire de Physique
Th\'eorique de l'Ecole Normale Sup\'erieure, 24 Rue Lhomond 75231 Paris, France}
\author{Thierry Giamarchi}
\affiliation{DPMC, University of Geneva, 24 Quai Ernest-Ansermet,
CH-1211 Geneva, Switzerland }
\author{Pierre Le Doussal}
\affiliation{CNRS-Laboratoire de Physique
Th\'eorique de l'Ecole Normale Sup\'erieure, 24 Rue Lhomond 75231 Paris, France}
\date{\today}
\begin{abstract}
We study the thermodynamics of disordered elastic systems, applied
to vortex lattices in the Bragg glass
phase. Using the replica variational method we compute the
specific heat of pinned vortons in the classical limit. We find
that the contribution of disorder is positive, linear at low
temperature, and exhibits a maximum. It is found to be important
compared to other contributions, e.g. core electrons, mean field and
non linear elasticity that we evaluate. The contribution of droplets is subdominant at weak disorder
in $d=3$.
\end{abstract}
\pacs{}
\maketitle

Understanding the temperature dependence of the
specific heat in glasses remains puzzling,
e.g. the linear low $T$ behaviour observed
in structural glasses \cite{zeller_chalspe_struct_glasses}
and spin glasses \cite{binder_spinglass_review}. In some
systems the crossover temperature from quantum to
classical behaviour may be quite low.
The phenomenological two levels system model \cite{anderson_twolevels}
yields a linear behaviour {\it both} in the classical
and quantum regime.
Although the classical
problem appears simpler there are only few, and
mostly mean field, solvable models of glassy systems
where one can actually compute the specific heat
\cite{binder_spinglass_review,derrida_rem}.
Ordered systems with continous symmetry admit
spin wave type excitations which yield a $T$-independent
specific heat $C_v=C_{eq}$ from the equipartition of the
energy. Non linearities such as quenched
disorder will cause a deviation which
is interesting to characterize and compare with the
linear contribution from the two level system
arguments.

A class of glassy systems recently much studied
are disordered elastic systems, ranging from vortex lattices
\cite{blatter_vortex_review,nattermann_vortex_review,giamarchi_book_young,giamarchi_vortex_review},
electron crystals \cite{giamarchi_wigner_review}, charge and spin
density waves \cite{gruner_book_cdw} to
disordered liquid crystals \cite{saunders_smectic_bragg_glass}.
In all of these systems the competition between
disorder and elasticity leads to pinning and glassy
behaviour. Specific heat measurements in density waves gapped systems
showed intermediate linear and sublinear regimes with
non-equilibration effects
 \cite{odin_chalspe_blue_bronze,lasjaunias_chalspe_noneq_sdw}.
In superconductors in a field, the $H$ and $T$ dependence of $C_v$
relates both to the symmetry of the order parameter
and to the thermodynamics of the vortex lattice.
If the contribution of the normal electrons in
the vortex cores dominate, the standard expectation is
that $C_v$ is linear in $T$ with a linear in $H$
dependence for $s$ wave \cite{caroli_core_states} and
$H^{1/2}$ for $d$-wave \cite{volovik_chalspe_dwaves}.
A specific heat linear in temperature has indeed been
measured in various materials. A
$H^{1/2}$-dependence has been observed and
argued for $d$-wave superconductivity in YbaCuO
\cite{moler_chalspe_ybacuo,wright_chalspe_ybacuo,revaz_chalspe_ybacuo}.
However the nonlinear $H$ dependence observed in other, a priori
non $d$-wave materials, is a well known puzzle, as discussed in
\cite{izawa_chalspe_borocabide,walti_chalspe_ube13}.
On the other hand, the contribution of
the phonons of the vortex lattice VL, the so-called ''vortons'', seems to be
within experimental resolution \cite{moler_fetter_kap}, and may lead, within the full temperature
range below VL melting, to more complex behaviors. These were
analyzed in absence of disorder, assuming
a dissipative quantum dynamics with friction $\eta$ arising from interactions between
vortons and vortex core electrons
\cite{blatter_chalspe,bulaevskii_chalspe}. It yields again
$C_v \propto \eta T$ with different $H$ dependence
but only for $T < T^D_v$, the vorton Debye-like temperature
which is poorly known (estimates in YBaCuO range from
well below 1K up to 10K in the superclean limit \cite{blatter_chalspe}).
Above $T^D_v$ one recovers the equipartition value to which the
specific heat anomaly was compared at melting
\cite{schilling_melting,zeldov_melting_95,roulin_chalspe_ybacuo,klein_private_mgb2}.
These analysis however neglect disorder and other non linearities
which for $T > T^D_v$ can be treated classically.

In this Letter we compute the specific heat of an elastic system
in presence of pinning disorder in the classical regime.
We show that disorder produces a substantial
rise above equipartition, linear at low temperature
and exhibiting a maximum at a characteristic depinning
temperature. We show that in $d > 2$ the contribution from the two well
droplet arguments is subdominant at weak disorder.
We find that the disorder contribution is
quite sizable compared to other contributions, e.g.
of the non linear elasticity that we also evaluate.
These results hold for a periodic
object, i.e. a Bragg glass, or for interfaces with continuous
degrees of freedom. In a companion paper \cite{schehr_chalspe_quantique} an analysis of the
quantum regime revealed, in the absence of dissipation,
a $C_v \sim T^3$ behaviour.

An elastic system, such as the vortex lattice (VL) with external
field aligned with $z$ axis, is described by a
$N$-component vector displacement field
$u_\alpha(R_i,z)$ ($N=2$ for the VL in $d=3$). The equilibrium positions $R_i$
form a perfect $N$-dimensional lattice of spacing $a$.
Interactions result in an elastic energy $H_{el}$ associated to
the phonons of the vortex lattices 
$ H_{el}[u] = \frac{1}{2} \int_{q} u_{\alpha}(q)
\Phi_{\alpha \beta}(q) u_{\beta}(-q)$. 
Here $\int_{q} \equiv \int_{BZ} \frac{d^2 q_\perp}{(2 \pi)^2}
\int_{-\pi/s}^{\pi/s}
\frac{d q_z}{(2 \pi)}$ denotes integration
on the first Brillouin zone, $q=(q_\perp,q_z)$ a $d$ dim-vector,
and $s$ is the distance between layers. For
the triangular VL:
\begin{equation}
\Phi(q) = (c_{66} q_\perp^2 + c_{44} q_z^2) P^T(q_\perp)
+ (c_{11} q_\perp^2 + c_{44} q_z^2) P^L(q_\perp) \label{el_tensor}
\end{equation}
with $P^L_{\alpha,\beta}(k)=k_\alpha k_\beta/k^2$ and
$P^T=\mathbb{I}-P^L$. The dispersion of elastic moduli is
implicit whenever needed. Impurity disorder is modeled by a short
range gaussian random potential with in plane correlator
$\Delta(r) = \delta s \epsilon_0^2 e^{-r^2/(4 r_f^2)}$ interacting with the
local vortex density. Here $2 r_f = \xi$ the superconducting
coherence length, $\epsilon_0 = (\Phi_0/4 \pi \lambda)^2$ is the vortex energy
scale per unit length along $z$ and $\delta$ a (small) dimensionless
disorder parameter \footnote{$\delta=\delta_\alpha \xi/s$ in the notations of
\cite{blatter_vortex_review}.}. The equilibrium Bragg glass phase (absence of
dislocations, $a/R_a \ll 1$, $R_a$ being the translational
correlation length) is described by the replicated partition
function $\overline{Z^n}= Tr e^{- \beta H_{\text{eff}}}$,
$\beta=1/T$ and $\overline {...}$ denotes disorder average. After
standard manipulations \cite{giamarchi_vortex_long} the replicated Hamiltonian becomes
$H_{\text{eff}}[u] = \sum_a H_{el}[u^a] + H_{dis}[u]$ with:
\begin{eqnarray}
&& H_{dis}= -  \frac{\beta}{2} \int d^2 r dz \sum_{ab} R(u_a(r,z) -
u_b(r,z)) \label{Hreplique}
\end{eqnarray}
Here $R(u)=\rho_0^2 \sum_K \Delta_K \cos(K \cdot u)$ in terms of
$\rho_0$ the average vortex density and the disorder harmonics
$\Delta_K = \int d^N u e^{i K \cdot u} \Delta(u)$ at the
reciprocal lattice vectors. More generally, an elastic manifold
(such as a directed polymer $d=1$) in a $N$-dimensional embedding
space in presence of a random potential $W(u,x)$ is described by a
similar model with $\overline{W(u,x) W(u',x)}=\delta^{(d)}(x-x')
R(u-u')$, $R(u)= - N V(u^2/N)$. We compute the specific heat
{\it per unit volume} $C_v(T) = - \frac{T}{\Omega}
\frac{\partial^2 \overline{F}}{\partial T^2}$ where $\overline{F}$
is the free energy, $\Omega=L^d$ the volume. We present the method using an isotropic
disorder and elasticity tensor $\Phi_{\alpha \beta}(q) = c q^2
\delta_{\alpha \beta}$, and generalize to the vortex lattice
later.

To obtain the low $T$ behavior, a first approach would be to
assume a {\it single minimum} of the energy $H[u]$ and expand
around it. One then finds:
\begin{eqnarray} C_v(T) =  C_{eq} + A T + O(T^2)
\label{low_T_chalspe}
\end{eqnarray}
The exact expression for the linear term $A=A_{1min}$, given in
\cite{schehr_chalspe_long}, involves cubic and quartic
anharmonicity in $H[u]$ in a given disorder realization. Disorder
averaging is only easy to perform perturbatively in disorder, yielding:
\begin{eqnarray}
A = A_{1min,pert} = - \frac{1}{6} J_1^3 (\nabla_u^2)^3 R(u)|_{u=0}
\label{pert}
\end{eqnarray}
with $J_1= \int_q 1/(c q^2)$ (for the manifold problem it
gives $A = 4 J_1^3 (N^2 + 6 N + 8) V'''(0)/(3 N))$. Although such a
single minimum low $T$ expansion is useful for pure systems, such
as the Sine-Gordon model, for disordered systems more than one
minimum typically exists beyond the Larkin length $R_c$. The
resulting contribution to the specific heat can then be estimated
combining the droplet picture with the two levels argument
\cite{anderson_twolevels,fisher_huse_droplets}. At each length
scale $l$ each of the subsystems $i \in (L/l)^d$ may have a low
lying secondary minimum (droplet) at excitation energy $E_i$
independently distributed with probability $P(E) dE =
\frac{dE}{E_c} (\frac{R_c}{l})^\theta {\cal F}(\frac{E
R_c^\theta}{E_c l^\theta})$ where $E_c$ is the typical pinning
energy $E_c=c r_f^2 R_c^{d-2}$, $\theta$ the free energy exponent.
Approximating the contribution
from scale $l$ to the specific heat as:
\begin{displaymath}
C_l = L^{-d} \sum_{i=1}^{(L/l)^d} (\frac{E_i}{T})^2 \frac{e^{-
\beta E_i}}{(1 + e^{- \beta E_i})^2} \approx \frac{\pi^2}{6}
\frac{T {\cal F}(0)}{E_c l^d}(\frac{R_c}{l})^\theta
\end{displaymath}
treating as independent two level systems, $C_v =
\int_{l>R_c} \frac{dl}{l} C_l$ is dominated by the smallest scales, yielding:
\begin{eqnarray}
A_{drop} \approx \frac{\pi^2 {\cal F}(0)}{6 (d+\theta)} E_c^{-1}
R_c^{-d} \label{drops}
\end{eqnarray}

While the two wells-droplet argument estimates as $A_{drop}$ the
contribution only of scales larger than $R_c$, one can only hope
to use $A_{1min}$ to estimate the contributions of scales smaller
than $R_c$. It is thus instructive to compare them. The
perturbative expression (\ref{pert}) is infrared divergent for $d
\leq 2$ as thermal fluctuations diverge. If one restricts by hand
the integral in (\ref{pert}) to $q >1/R_c$ one finds a
contribution of the same order than the droplet one $A_{drop}$. In
$d > 2$ the integral is instead controlled by small scales and the
droplet contribution is then {\it subdominant}. The two levels droplet
model of \cite{anderson_twolevels} can be improved by including
anharmonicity in each well, for identical wells it simply adds
$A=A_{1min} + A_{drop}$.

The variational method
\cite{mezard_variational_replica,giamarchi_vortex_long} extends
these phenomenological considerations into a first principle
quantitative calculation in which the Larkin length naturally
appears in a self consistent way. We introduce a Gaussian trial
Hamiltonian $H_{0} = \frac{1}{2} \int_q G^{-1}_{\alpha
\beta,ab}(q) u_{\alpha}^a(q) u_{\beta}^b(-q)$ which minimizes the
variational free energy $F_{\text{var}} = F_0 + \langle
H_{\text{eff}} - H_{0} \rangle_{H_0}$, where $F_0$ denotes the
free energy calculated with $H_0$. The specific heat $C_v =
\lim_{n \to 0} -\partial_T
\partial_{\beta} \frac{\overline {Z^n}}{n \Omega}$ can be reexpressed as:
\begin{eqnarray}
C_v = \partial_T  \lim_{n \to 0} \frac{1}{n \Omega} ( \langle
\sum_a H_{el}[u_a] \rangle_{H_{\text{eff}}}  + 2\langle
H_{dis}[u]\rangle_{H_{\text{eff}}} ) \label{expr_chalspe}
\end{eqnarray}
Note the factor of $2$ due to the $\beta$ dependence of the disorder
term.
Here we evaluate these averages using the variational hamiltonian $H_0$
instead of the exact one $H_{\text{eff}}$. Thanks to the variational equations
this is equivalent to
$C_v(T) = -(T/\Omega) \partial^2 F_{\text{var}}/\partial T^2$

We applied this variational approach to some pure models and
checked that it is quite accurate \cite{schehr_chalspe_long}. At
low $T$ it exactly matches the expansion around the minimum
$A_{var}=A_{1min}$ and we checked, for the Sine-Gordon
model in $d=2$, that it is identical to the low $T$ expansion of the exact
result \cite{papa_sinegordon_exact}.

\begin{figure}
\centerline{\includegraphics[width=\figwidth]{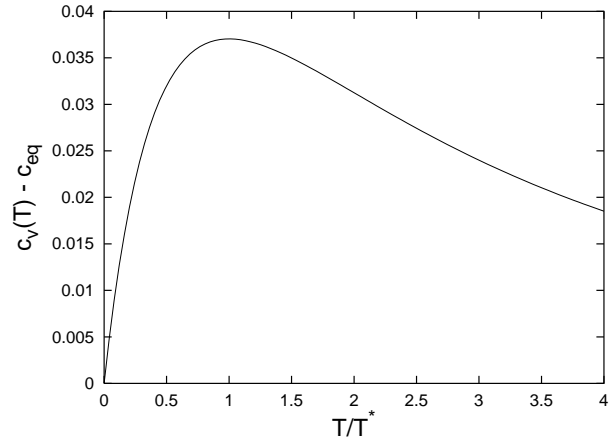}}
\caption{Specific heat $C_v(T) - C_{eq}$ in arbitrary units as a
function of $T/T^*$.} \label{plot_chalspe}
\end{figure}

In the disordered case the solution of the variational equations
requires Replica Symmetry Breaking (RSB). One
denotes $\tilde{G}(q) = G_{aa}(q)$ and parametrizes $G_{a \neq
b}(q)$ by $G(q,u)$, where $0 < u <1$ and similarly
for ${\cal B}_{ab}(x=0) = {\cal B}_{ab} = \langle [u_{\alpha}^a(x) -
u_{\alpha}^b(x)]^2\rangle/N$ with $\tilde{{\cal B}}
= 0$ and ${\cal B}(u)$. For $d>2$ one finds
\cite{mezard_variational_replica,giamarchi_vortex_long}
a continuous RSB with a breakpoint $u_c$ and
\begin{eqnarray}
&& {\cal B}(u > u_c) = {\cal B} =  2 \gamma T J_1(\Sigma) \label{break_point}\\
&& 1 = - 4 \gamma \hat{V}''({\cal B}) J_2(\Sigma) \label{marginality} \\
&& J_n(z) = \int_q \frac{1}{(cq^2 + z)^n}
\quad , \quad \Sigma = c R_c^{-2}  \label{larkin_length}
\end{eqnarray}
where $\gamma=1$, $\hat{V}({\cal B}) = - \frac{1}{N} \rho_0^2 \sum_K
\Delta_K \exp(-
{\cal B} K^2/2)$ and $R_c$ is the Larkin length \footnote{for the manifold
$\hat{V}({\cal B})$ is defined from $V({\cal B})$ in \cite{mezard_variational_replica}}. Eq. (\ref{marginality}) is
the so called marginality condition (MC) which also holds for the
one step solution in $d=2$. Starting from the expression:
\begin{equation}
\frac{1}{N\Omega} \overline{\langle H \rangle} = \int_{q}
\frac{1}{2} q^2 \tilde{G}(q) + \frac{1}{T} \int_0^1 du [\hat{V}(0) -
\hat{V}({\cal B}(u))]
 \label{Hquantmoy}
\end{equation}
which, as $B(w)={\cal B}(u)$ (setting $w=u/T$) and $w_c=u_c/T$
\footnote{one finds $w_c =
4 \hat{V}'''({\cal B})  J_2(\Sigma)^3/J_3(\Sigma)$}, turns out to
depend implicitly on $T$ only through $\Sigma$ and ${\cal B}$, and using
(\ref{break_point}-\ref{larkin_length}) and (\ref{expr_chalspe}),
one obtains for the specific heat
\begin{eqnarray}
&& C_v(T) = C_{eq} + \frac{N}{T^2} F({\cal B})
\label{chaleur_spe_classique}  \\
&& F({\cal B}) = \hat{V}({\cal B}) - \hat{V}(0) - {\cal B}
\hat{V}'({\cal B}) + \frac{1}{2} {\cal B}^2 \hat{V}''({\cal B}) \nonumber
\end{eqnarray}
where $C_{eq}=\frac{1}{2} {\cal N}/\Omega=\frac{N}{2}\int_q$ is
the equipartition value, i.e. half the total number of modes per
unit volume \footnote{$C_{eq}$ doubles upon adding the kinetic energy}.
Eq. (\ref{chaleur_spe_classique}) is valid
for any $T$, for periodic objects (Bragg glass) as well
as manifolds, and independently of the replica structure of the
solution (if broken, provided MC holds). One thus
finds that disorder increases the specific heat which has now a
maximum and decreases back to equipartition $C_{eq}$ at high $T$.
Expanding (\ref{chaleur_spe_classique}) at low $T$ one finds again
(\ref{low_T_chalspe}) with an amplitude $A_{var} = \frac{8}{3!} N
\hat{V}'''(0) J_1(\Sigma_{T=0})^3$. For weak disorder $R_c>a$, dispersionless
elasticity and $r_f<a$, one finds from (\ref{break_point}-\ref{larkin_length})
$A_{var} \approx \tilde{A}/(c r_f^2 R_c a^3)$, $\tilde{A}=\frac{\pi
N(N+4)}{96}$
indeed larger by a factor $(R_c/a)^3$ than (\ref{drops}). It also
confirms the above discussion:
thanks to RSB (i.e. $\Sigma
\neq 0$), the problems (e.g. in $d=2$) of single minimum (i.e.
replica symmetric) perturbation theory are cured, the Larkin
length being the natural scale. A plot of $C_v(T)$ is
shown on Fig.~\ref{plot_chalspe}

The generalization to the vortex lattice using (\ref{el_tensor})
and $N=2$ is straightforward. Eq.
(\ref{chaleur_spe_classique}) still holds, but in the formulaes
(\ref{break_point},\ref{marginality}) which determine ${\cal B}$ and
$\Sigma=c_{66} R_c^{-2}$ one sets $\gamma=\frac{1}{2}$ and replaces
$J_n(\Sigma)=
\int_q (c_{66} q_{\perp}^2 + c_{44} q_z^2 + \Sigma)^{-n}$ (we
neglect compression modes, i.e. $c_{66}/c_{11} \ll 1$).

Taking
dispersion in $c_{44}$ and anisotropy into account yields several regimes
analyzed in \cite{schehr_chalspe_long}, and for simplicity we present
here only results for the weak disorder regime $R_c > a$.
The fluctuation of a vortex position is then measured from ${\cal B} = T
J_1(\Sigma) \approx T J_1(0) = c_L^2 a^2 T/T_m$, where
$T_m$ is the melting temperature and $c_L \approx 0.1-0.2$ is the
Lindemann number. Inserting in (\ref{chaleur_spe_classique}) this
yields $C_v(T)$ which has a maximum for $T=T^*$ i.e. ${\cal B}={\cal B}^*$
determined by solving $4 F({\cal B}^*)=({\cal B}^*)^3 \hat{V}'''({\cal
B}^*)$. These
fluctuations are estimated as $J_1(\Sigma) \approx 1/(4 a
\sqrt{c_{66} c_{44}})$ where here and below we denote $c_{44}=
c_{44}(q_\perp=\pi/a)$ (below the dimensional crossover field,
$c_{66}/(c_{44} a^2) \ll 1/s^2$, the integral over $q_z$ can be
extended up to infinity and the integral is dominated by
$q_\perp=\pi/a$).

We can now discuss in details the behaviour of $C_v(T)$ for $r_f <
a$. In this limit $\hat V({\cal B})$ takes the dependence $\hat
V({\cal B}) \approx - D/(2
r_f^2 + {\cal B})$, with $D=\delta r_f^2 s \epsilon_0^2/a^2$,
for ${\cal B} \ll a^2$ which holds up to melting. One finds:
\begin{eqnarray}
&& C_v(T) = C_{eq} + A T/(1 + T/(2 T^*))^3  \label{cvortex} \\
&& A \approx \frac{\delta}{128} c_L^2 (\frac{\epsilon_0^2}{c_{66} c_{44} a^4})
\frac{s a^2}{r_f^6 T_m}
\end{eqnarray}
The maximum occurs for ${\cal B}^*=r_f^2$ and thus $T=T^* = T_m
r_f^2/(c_L^2 a^2)$, i.e. the so called depinning temperature
\cite{blatter_vortex_review},
which can be below melting. The amplitude $A$ (and large $T$ tail)
is independent of the detailed form of $\hat V({\cal B})$ and is strongly
enhanced by the anisotropy and dispersion of $c_{44}$. The value
at the maximum is $C_v^{*}-C_{eq}=8 A T^*/27$. Using
$c_{44} \approx \epsilon_0 \epsilon^2/a^2$, $c_{66} \approx \epsilon_0/(4 a^2)$
the calculation
of $J_2(\Sigma)$ shows \cite {schehr_chalspe_long} that $R_c >a$ holds
for $\delta < \delta_c =
4 \pi^{3/2} \frac{\epsilon r_f^4}{s a^3}$. At this
value of the disorder one obtains
$A \approx c_L^2/(T_m r_f^2 a \epsilon)$ and $C_v^{*}-C_{eq}
\approx 1/(a^3 \epsilon)$ (
$\epsilon<1$ the anisotropy parameter \cite{blatter_vortex_review}). $C_v^*-C_{eq}$ should be
compared with the equipartition value, estimated as $C_{eq} = 1/(s
a^2)$. These become comparable around the
dimensional crossover field $B=B_{cr}$ such that $s \sim a \epsilon$.


The above classical contribution (\ref{cvortex}) will hold only
above the vorton Debye temperature $T_v^D$, below which quantum
effects become important. $T_v^D$ depends on the vortex mass, the Hall
force and the
friction force, which arise from the coupling
of moving vortices with the normal electrons bath in presence of
scattering. Estimates for $T_v^D$ range \cite{blatter_chalspe}
from $10^{-3} T_c$ in dirty (friction dominated) materials to
$10^{-1} T_c$ in superclean limit (Hall dominated). There should thus
exist a broad regime of temperature and field where the Bragg glass is
stable and the result (\ref{cvortex}) holds.

To assess whether this contribution (\ref{cvortex}) is observable, let us compare
it with other terms linear in $T$ in $C_v$. First the normal
electrons in the vortex core lead to \cite{caroli_core_states}:
\begin{eqnarray}
&& {C_v}_{\text{core}}(T) \approx \frac{T}{T_f k_f^{-2} s}
(\frac{H}{H_{c2}})^\alpha
\end{eqnarray}
where $T_f$ is the Fermi temperature of the normal metal,
$\alpha=1$ for $s$-wave superconductor \footnote{for $T \gg \hbar
\omega_0$ the so-called mini-gap} and $\alpha=1/2$ for lines of
nodes in the gap \cite{volovik_chalspe_dwaves}. Given the large
ratio $T_f/T_m = O(10^2)$ for e.g. YBaCuO, one finds, comparing
$A_{core}$ and $A_{dis}$ that the contribution from the cores can
be comparable or smaller than the one from the disorder.

There are other contributions from the vortex lattice as well. The
mean field specific heat \cite{bulaevskii_chalspe} being $C_{mf}
\approx \epsilon_0 T/(T_c^2 a^2)$, the ratio $A_{dis}/A_{mf}
\approx (c_L^2 a T_c/(r_f T_m))^2$ can be large. We have also
computed the contribution from non linear elasticity of the VL.
Performing, as in \cite{dodgson_anharmonicity}, an
expansion of the vortex interaction energy $\int dz
\sum_{ij}\epsilon_0 K_0((R_i - R_j + u_i(z) - u_j(z))/\lambda)$,
we obtain:
\begin{equation}
A_{nl} = \rho_0 \epsilon_0 a^{-4} (\frac{c_L^2 a^2}{T_m})^2 (
\gamma_4 + \gamma_3 \sqrt{c_{44}}{c_{66}} (\frac{c_L^2 a^2}{T_m})
\frac{\epsilon_0}{a} )
\end{equation}
up to $O((a/\lambda)^2)$ where the numerical prefactors $\gamma_4$
and $\gamma_3$ are complicated lattice sums, given in
\cite{schehr_chalspe_long}. One can then estimate $A_{nl}/A_{dis}
\sim r_f^2/a^2$ and this contribution is likely to remain small until
melting. None of these contributions \footnote{we can also neglect $T$ dependence
of $V$ from $\lambda$ and $\xi$ away from $H_{c2}$} is expected to
exhibit a maximum at scale distinct from $T_m$.

To conclude, we found by explicit calculation that the vortex
lattice classical contribution to $C_v$ due to disorder can be
important at the very least in the range $10^{-1} T_c$ to $T_m$,
and possibly more.  The coefficient $A_{dis}$ of the low $T$ linear
behavior is magnetic field dependent, $A_{dis} \propto 1/(T_m a)$.
It is convenient to express results using
the melting temperature $T_m$, which is experimentally measured
\cite{schilling_melting,zeldov_melting_95,roulin_chalspe_ybacuo}
and can also be estimated from the standard elastic expression of $T_m = 4a^3
\sqrt{c_{66}c_{44}}c_L^2$ allowing to extract directly the magnetic field
dependence. $T_m \sim 1/\sqrt{B}$ leads to $T^* \propto \sqrt{B}$
and $A_{dis} \propto B$. $C_v(T)$ exhibits, as compared to other
contributions, a distinct maximum around the depinning temperature
scale, whenever smaller than melting. A crude estimate of $T^*$ is
$T^* \simeq T_m B/(c_L^2 H_{c2})$, which for a typical $c_L \sim
0.12$ gives a $T^*$ which is a fraction of $T_m$ for fields of
about a Tesla (a fraction of order unity
for YBaCuO around $B \simeq 10T$ the so-called tricritical point
\cite{roulin_chalspe_ybacuo}).
It would be very interesting to perform precise
measurements of $C_v$ to check the present proposal.

\begin{acknowledgments}
We thank T. Klein, P. Monceau, A. Junot and C.M. Varma for useful discussions.

\end{acknowledgments}


\end{document}